\documentclass[12pt]{iopart}
\usepackage{iopams}
\usepackage{cite}
\usepackage{graphicx}

\begin{document}
\title[Flashing subdiffusive ratchets]{Flashing subdiffusive ratchets in viscoelastic media}

\author{Vasyl Kharchenko$^{1,2}$, Igor Goychuk$^1$}

\address{$^1$ Institute of Physics, University of Augsburg, Universit\"{a}tstr. 1, D-86135 Augsburg, Germany}
\ead{vasyl.kharchenko@physik.uni-augsburg.de}
\address{$^2$ Institute of Applied Physics, 58 Petropavlovskaya Str., 40030 Sumy, Ukraine}
\ead{igor.goychuk@physik.uni-augsburg.de}

\begin{abstract}
We study subdiffusive ratchet transport in periodically and randomly flashing potentials. Central 
Brownian particle is elastically coupled to surrounding auxiliary Brownian quasi-particles 
which account for the influence 
of viscoelastic environment.  Similar to standard dynamical modeling of Brownian motion,  the 
external force influences only the motion of central particle not affecting directly 
the environmental degrees of freedom (see video). Just a handful of
auxiliary Brownian particles suffice to model subdiffusion over many temporal 
decades. Time-modulation of the potential violates the symmetry of thermal detailed balance
and induces anomalous subdiffusive current which exhibits a remarkable quality at low temperatures,
as well as a number of other surprising features such as saturation at low temperatures, 
and multiple inversions of the transport direction upon a change of the 
driving frequency in nonadiabatic regime. Our study generalizes classical Brownian motors towards
operating in sticky viscoelastic environments like cytosol of biological cells or dense 
polymer solutions.

\end{abstract}

\pacs{05.40.-a, 05.10.Gg, 87.16.Uv}

\maketitle

\section{Introduction}

Physics of noise-assisted driven transport presents currently a well-established area of research
\cite{ReimannRev,MarchesoniRev}. Most papers are devoted to classical transport in neglecting 
non-Markovian and even inertial effects.  
The corresponding stochastic nonlinear dynamics can be described by a Langevin equation in 
time-dependent potentials and/or by associated Fokker-Planck equation for the noise-averaged dynamics
of the probability density of an ensemble of moving particles. Key ingredients are 
nonlinear dynamics in periodic potentials unbiased on average, friction and thermal noise related by
the fluctuation-dissipation theorem (FDT), and an external driving which violates the symmetry of thermal
detailed balance ensured by the FDT at thermal equilibrium. The emerged dissipative out-of-equilibrium 
directed transport is necessarily accompanied  by an entropy production and related heat dissipation.
% and requires a source of free energy to persist. 

Such profoundly out-of-equilibrium transport should be distinguished from other possibilities such
as transport in a running potential, where the particle remains bound to a potential well which moves
in space. Basically, in this later case one does not need even a periodic potential. Any trapping confining
potential $U(x)$ with a deep minimum at $x=x_0$ will convey a transport if this minimum is moving with  
velocity $v$, i.e. $U(x)$ is replaced by $U(x-vt)$.  Such a ``peristaltic'' 
transport is clearly not related in essence to any entropy
production and for a periodic running potential it can be symbolized by the Archimedean screw pump. This
is a purely mechanical system, where both the friction and the noise are not principal for the transport occurrence.
Quantum-mechanical counterparts of the Archimedean pump are also well-known \cite{Switkes}.
A peristaltic pump can be also realized with highly overdamped isothermic thermodynamic systems operating 
close to thermal 
equilibrium, where dissipation keeps the particle fluctuating near to the potential minimum, and 
the energy lost due to friction is perpetually restored due to stochastic impact of 
the environment -- the physical content of FDT. Friction plays here in fact a constructive role. 
Moreover, neglect of the energy gain from hot environment, at odds
with FDT, is one of the common mistakes in the literature, leading to erroneous belief that one must always 
minimize dissipation and cool the environment in order 
to achieve at the highest efficiencies of isothermal engines possible. This is not necessarily so.
For example, the
so-called dissipationless Hamiltonian ratchets \cite{Pseudo} cannot do any useful work at all 
against a load, i.e. are pseudo-ratchets 
with zero efficiency.
A dissipative system can remain locally close to thermal equilibrium,  but the form of the 
potential is cyclically and adiabatically slow changed by a driving force 
so that excess, uncompensated  heat exchange 
between the particle and the thermal reservoir can be minimized.
In this quasi-equilibrium scenario, the work required to change the form of potential 
(to move the minimum) can be transformed with minimal heat losses into the work against the load which 
opposes this potential modulation. This is why
such an \textit{isothermal} machine considered as a free energy transducer 
can in principle operate with the efficiency, defined as 
the ratio of the work against a load to the free energy spent to drive a working cycle, 
close to one \cite{Hill,Tsong,Chen,Astumian}.
For example, the efficiencies of highly optimized biological ionic pumps as high as $0.75$  
are common \cite{NelsonBook} and the efficiency of ATP synthase can approach the theoretical
maximum of one \cite{Kinosita,Toyabe,Romanovsky}.

\begin{figure}
 \centering
 \includegraphics[width=80mm]{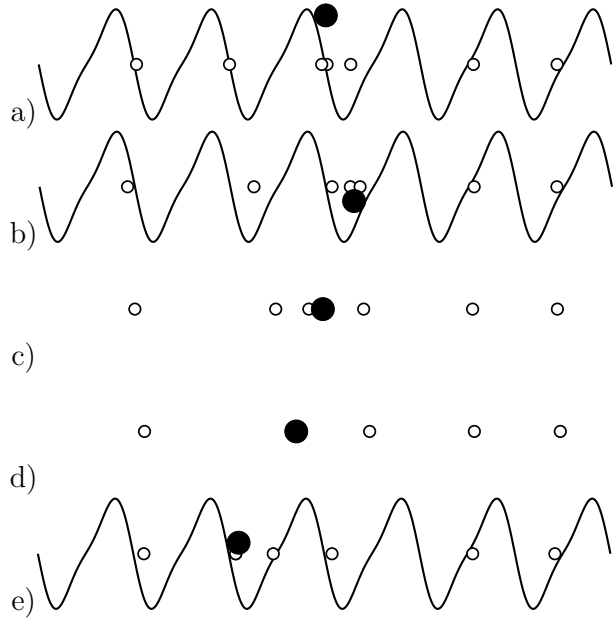}
 \caption{Snapshots of numerical simulations of the dynamics in Eq. (\ref{embedding1})
  at different instants of time  a) $t=0.1$, b) $t=0.4$, c) $t=1.6$, d) $t=2.0$, e) $t=2.5$.
 One Brownian particle (big filled circle) is
 coupled to auxiliary particles (small empty circles) at temperature $T=0.1$ and moves in 
 a periodically flashing with 
 frequency $\nu=1.0$
potential (\ref{U(x)}) with amplitude $U_0=0.75$.}
 \label{Fig1} 
\end{figure}

The primary focus of this paper is different, on the thermal noise assisted transport, where the thermal noise
is assumed to play a profound and constructive role. 
A paradigmatic example  is provided here by transport in flashing potentials \cite{Ajdari}, 
such as one in  figure \ref{Fig1}. 
In a standard Markovian overdamped setup, 
when the potential is off, an initially localized particle
diffuses with the position variance growing linearly, $\langle \delta x^2(t)\rangle \propto t$,
which corresponds to normal diffusion.
When the potential is on, the particle relaxes to a minimum of the potential and the probability
distribution becomes at least bimodal with a larger peak corresponding to sliding down a less steep
side of the potential (as a larger basin of attraction corresponds to a larger distance between the minimum 
and maximum of a spatially asymmetric but periodic potential).
When flashing repeats, either periodically or stochastically, but sufficiently
slow, the net transport is expected to emerge in the left, natural direction, with the
averaged particle's position growing linearly in time. Such normal transport is characterized
by  mean velocity. This natural direction is opposite to one
 in the fluctuating tilt potential ratchets of the same potential form \cite{Magnasco,Bartussek}.
For a fixed unbiased on average 
potential, the total heat
exchange between the particle and its environment is zero, FDT holds, and  directed transport is forbidden
by the symmetry of thermal detailed balance.
Out-of-equilibrium potential fluctuations  violate this symmetry  and induce directed transport.
There emerges an overall uncompensated excess heat flow to the environment associated with the corresponding
entropy production. A part of the energy put in the repeating potential flashing is dissipated as excess 
heat and a part can be used to do useful work against a load.
If the load is absent all the consumed energy is dissipated as excess heat (futile motor) because the 
mechanical energy of the motor particle remains on average not changed. This is a well-established by now
physical picture \cite{Julicher}. This basic model explains e.g. operating of single-headed kinesin motors 
 \cite{NelsonBook}, where the energy to drive stochastic cycles is provided in effect by the energy of 
ATP hydrolysis.

Generalization of this approach to account for a non-Markovian friction with memory is not trivial.
The memory effects appear e.g. due to viscoelasticity of the environment \cite{Maxwell,Gemant,Mason,Jones,Waigh,
G09,G12}.  
Even if a corresponding 
Generalized Langevin Equation (GLE) \cite{GLE,GLE2,LEBook,WeissBook} is well known for any 
linear model of friction with the memory, the
corresponding non-Markovian Fokker-Planck equation (NMFPE) \cite{Adelman,Hanggi78,Hanggi82,Hynes,GH07} 
remains simply unknown 
for general nonlinear force-fields.
The cases of constant force, or a linear in coordinate force, where the corresponding NMPPEs are known 
\cite{Adelman}, 
are not especially useful in the present context. Especially interesting is the case of a power law decaying
memory kernel \cite{Gemant} which corresponds to subdiffusion and is associated with the Cole-Cole dielectric response
of viscoelastic media \cite{Cole,G07}. 
Such viscoelastic subdiffusion has recently been found relevant also for transport 
in polymer networks \cite{Amblard,Caspi}, cytosol of biological cells \cite{Guigas} and akin 
crowded fluids \cite{Szymanski}. 
The corresponding fluctuating tilt or rocking ratchets in viscoelastic media have  recently been proposed
and studied in Refs. \cite{G10,arxiv} with a number of quite unexpected and surprising properties revealed.
The interest to ratchet effect in glass-like environments is growing  \cite{Giacomo}. 
Flashing ratchet transport in dynamically disordered potentials was proposed and studied
earlier in Ref.\cite{Harms}. Transport in disordered potentials is known to be equivalent within
the mean field effective medium approximation to a continuos time random walk (CTRW) approach with 
independent residence
times in traps \cite{Hughes,Bouchaud,Metzler}.
This places our
research into a general context of anomalous transport processes in complex media 
\cite{Hughes,Bouchaud,Metzler,WSBook}. However, our approach to anomalous transport
in viscoelastic media
is very different from one based on uncorrelated CTRW. 
% as reviewed in Ref. \cite{}. 
This work is devoted to flashing viscoelastic ratchets  which turn out
to be no less surprising than rocking viscoelastic ratchets \cite{G10,arxiv} 
and we shall take the advantage of the approach to anomalous subdiffusive transport
developed recently in Refs. \cite{G09,G10,G12}.

\section{Dynamical approach to viscoelastic transport and stochastic modeling}

We start with a well-established dynamical approach to the theory of Brownian motion \cite{GLE2,WeissBook}. 
The environment
is modeled by a set of $N_0$ particles with masses $m_i$ harmonically coupled with the spring constants $\kappa_i$ 
to the Brownian particle of mass $m$: 
\begin{eqnarray}
m\ddot x & = & f(x,t)- \sum_{i=1}^{N_0} \kappa_i (x-q_i), \label{A2}\\
m_i \ddot q_i & = & \kappa_i (x-q_i)\;, \label{A3}
\end{eqnarray}
where $q_i$ are the coordinates of the environmental particles, $x$ is the coordinate of the Brownian
particle, and $f(x,t)$ is an external force which affects only the motion of Brownian particle but not
the environment.  
As usually, using the Green function of harmonic oscillator one can express the bath oscillators 
coordinates $q_i(t)$ in Eq. (\ref{A3}) via the initial values $q_i(0)$ and $p_i(0)$ for
arbitrary $x(t)$:
\begin{eqnarray} \label{A5}
q_i(t)& = &q_i(0)\cos(\omega_i t)+\frac{p_i(0)}{m_i\omega_i}\sin(\omega_i t)+
\frac{\kappa_i}{m_i\omega_i}\int_0^t \sin[\omega_i(t-t')]x(t')dt' \nonumber \\
& = & [q_i(0)-x(0)]\cos(\omega_i t)+\frac{p_i(0)}{m_i\omega_i}\sin(\omega_i t) 
+x(t)\nonumber \\
& & -\int_0^t\cos[\omega_i(t-t')]\dot x(t')dt\;. 
\end{eqnarray}
Here, $\omega_i=\sqrt{\kappa_i/m_i}$ are the bath oscillators frequencies 
and the integration by parts has
been used to obtain the second equality. By substituting (\ref{A5}) into (\ref{A2}) one immediately
obtains
\begin{equation}\label{GLE}
m\ddot x+\int_{0}^{t}\eta(t-t')\dot x(t')dt'=f(x,t)+\xi(t),
\end{equation}
where
\begin{eqnarray}\label{kernel}
\eta(t)=\sum_{i}\kappa_i\cos(\omega_i t),
\end{eqnarray}
and
\begin{eqnarray}\label{noise}
\xi(t)=\sum_i \kappa_i \left ( [q_i(0)-x(0)]\cos(\omega_i t)+
\frac{p_i(0)}{m_i\omega_i}\sin(\omega_i t) \right ) \; .
\end{eqnarray} 
Eq. (\ref{GLE}) is still a purely dynamical equation of motion which is exact, the 
dynamics of irrelevant degrees of
freedom is but excluded. For example, it describes a time-reversible dynamics, if external field $f(x,t)$ does not 
violate the time-reversal symmetry\footnote{Time-reversal symmetry can be \textit{dynamically} 
broken by an external time-dependent 
field. 
For example, a harmonic mixing driving, $f(t)=A_1\cos(\Omega t)+A_2\cos(2\Omega t+\phi)$ does violate 
the time-reversal symmetry dynamically for $\phi\neq 0$ \cite{GH00}.}. Then the trajectories
governed by Eq. (\ref{GLE}) are also  most obviously  time-reversal symmetric (see  \cite{Chaos} on this 
point from stochastic perspective). 
This is contrary to a widespread 
misperception. 
The \textit{statistical} irreversibility comes about on the level of a bunch of trajectories \textit{averaged} 
over different initial realizations
of $q_i(0)$, $p_i(0)$. The loss of information due to averaging or coarse-graining is a well known source of 
irreversibility leading to statistical description of dynamical systems (see  e.g. \cite{textbook}). 
As a matter of fact, the dynamics described by Eq. (\ref{GLE}) is just a projection of a highly dimensional 
Hamiltonian dynamics onto two dimensional $(x,p)$ phase subspace.

Furthermore, one proceeds as in a typical molecular dynamics setup. The initial
positions $q_i(0)$ and momenta $p_i(0)$ of the environmental oscillators in Eq. (\ref{noise}) are sampled 
(this is the only non-dynamical element in the theory) from a 
canonical distribution at temperature $T$,
\begin{equation}
\hspace{-0.9cm}\rho(\{q_i(0),p_i(0)\}|x(0))=
\frac{1}{Z}\exp\left [
-\frac{1}{2k_BT}\sum_{i=1}^{N_0}\left( \frac{p_i^2(0)}{m_i}+\kappa_i[q_i(0)-x(0) ]^2 \right ) \right ],
\end{equation}
conditioned on the initial position of the Brownian particle $x(0)$, where $Z$ is the statistical sum
of bath oscillators. Here one assumes that initial velocities of the bath
oscillators are centered at zero, i.e. the medium is not moving as whole. Likewise, the average 
$\langle q_i(0)\rangle =x(0)$, i.e. the medium is initially equilibrated adjusting to the Brownian
particle localized at $x(0)$.
The corresponding random force $\xi(t)$
becomes a stochastic process which is obviously Gaussian and can be completely characterized by its first
two statistical moments. First moment is obviously zero, $\langle \xi(t)\rangle=0$. Furthermore,  with Gaussian averages 
$\langle p_i(0)p_j(0)\rangle=\delta_{ij}m_ik_BT $, $\langle x_i(0)x_j(0)\rangle=\delta_{ij}k_BT/\kappa_i $,
$\langle x_i(0)p_j(0)\rangle=0$ it is easy to show that    
\begin{equation}\label{FDR}
\langle \xi(t')\xi(t)\rangle =k_B T\eta(|t-t'|)
\end{equation} 
for any set of bath oscillators. This is the celebrated fluctuation-dissipation relation, or the second
fluctuation dissipation theorem by Kubo \cite{GLE}. The bath oscillators are conveniently characterized by the
spectral density \cite{WeissBook}
\begin{eqnarray}
J(\omega)=\frac{\pi}{2}\sum_i\frac{\kappa_i^2}{m_i\omega_i}\delta(\omega-\omega_i)=\frac{\pi}{2}\sum_i
m_i\omega_i^3\delta(\omega-\omega_i)\;.
\end{eqnarray}
It allows to express $\eta(t)$ as $\eta(t)=(2/\pi)\int_0^{\infty}d\omega J(\omega)\cos(\omega t)/\omega$
and the noise spectral density as 
$S(\omega)=2k_BT J(\omega)/\omega$ via  the Wiener-Khinchin theorem, $S(\omega)=
\int_{-\infty}^{\infty}\langle \xi(t)\xi(0)\rangle e^{i\omega t}dt$. The choice 
$J(\omega)=\eta_{\alpha}|\sin(\pi\alpha/2)|\omega^{\alpha}$ with 
$0<\alpha<2$, yields the fractional Gaussian noise (fGn) $\xi(t)$ introduced by  Mandelbrot
and van Ness \cite{Mandelbrot}. For $0<\alpha<1$ (sub-Ohmic thermal bath \cite{WeissBook}), 
\begin{equation}\label{PL}
\eta(t) = \eta_{\alpha}t^{-\alpha}/\Gamma(1-\alpha),
\end{equation} 
where $\Gamma(z)$ is a standard gamma function. This choice yields subdiffusion asymptotically, 
$\langle \delta x^2(t)\rangle \sim t^\alpha$,
for an ensemble of particles \cite{WeissBook}.
The corresponding GLE (\ref{GLE}) is termed also the fractional GLE, or FLE \cite{FLE,FLE2,GH07,G12} 
upon the use of the notion of fractional Caputo derivative to shorthand
the frictional term, and the coefficient $\eta_{\alpha}$ is termed then the fractional friction coefficient.
Such a dynamical 
modeling requires a very large $N_0\to\infty$ number of the bath oscillators with a quasi-dense spectrum.

\subsection{Stochastic modeling}

Alternatively, one can use just a handful of $N$ auxiliary Brownian particles clouding around 
the central particle \cite{GH11,G12} while modeling the rest as friction and noise
acting on these representative ones.   Such Brownian quasi-particles can serve to model sticky viscoelastic media.
A particle clouded by other particles reminds conceptually
polaron  picture in condensed matter physics.  
Then, one replaces
Eqs. (\ref{A2},\ref{A3}) with 
\begin{eqnarray}
 m\ddot x & = & f(x,t)-\sum_{i=1}^{N}k_i(x-x_i) \; \label{osc1}\\
 m_i\ddot x_i & = & k_i (x-x_i)-\eta_i\dot x_i+\sqrt{2\eta_ik_BT}\zeta_i(t) \label{osc2}\;,
\end{eqnarray}
where $x_i$ are the coordinates of auxiliary particles, $k_i$ are the corresponding coupling constants, 
$\sqrt{2\eta_ik_BT}\zeta_i(t)$ are thermal Gaussian forces, with zero range of correlation, 
$\langle \zeta_i(t)\zeta_j(t')\rangle=\delta_{ij}\delta(t-t')$, and $\eta_i$ are the corresponding friction
coefficients. Moreover, the overdamped limit for these auxiliary stochastic medium oscillators, $m_i\to 0$,
yields
\begin{eqnarray}\label{embedding1}
\dot x & = & v\;,\nonumber \\
m\dot v & = & f(x,t)+ \sum_{i=1}^{N}u_i(t) \; \\
\dot u_i& = &-k_i v-\nu_iu_i+\sqrt{2\nu_i k_i k_BT}\zeta_i(t) \;,\nonumber
 \end{eqnarray} 
where $\nu_i=k_i/\eta_i$ are the relaxation rates of the introduced viscoelastic forces $u_i=-k_i(x_i-x)$. 
The last equation for $u_i$ is similar to the Maxwell's relaxation equation for viscoelastic force in  
macroscopic theory of viscoelasticity \cite{Maxwell} which is augmented by the corresponding Langevin force 
in accordance with FDR. Such a description was introduced in Refs. \cite{G09,G10,GH11,G12} 
to model anomalous Brownian motion in complex viscoelastic media within a generalized Maxwell model.
For a particular case of the only one auxiliary particle (Maxwell model of viscoelasticity), the
earlier description in Refs. \cite{Marchesoni,Straub} leading to GLE with thermal Ornstein-Uhlenbeck noise is 
readily reproduced.
Excluding the dynamics of auxiliary variables $u_i(t)$ and assuming that 
initially forces $u_i(0)$ are thermally Gaussian-distributed  
with zero mean and dispersion $\langle u_i^2(0)\rangle=k_ik_B T$ yields again the GLE (\ref{GLE}) 
with the corresponding memory kernel  presented by a sum of exponentials,
\begin{equation}
 \eta(t)=\sum\limits_{i=1}^{N}k_i e^{-\nu_i t}.
\end{equation}
The corresponding noise $\xi(t)$ presents the sum of Ornstein-Uhlenbeck noises.

The power-law memory kernel (\ref{PL}) can be reliably  approximated by such a sum \cite{G09} over a large 
time-interval $[t_l,t_h]$
by choosing  $\nu_i=\nu_0/b^i$  and $k_i=C_{\alpha}(b)\eta_{\alpha}\nu_i^{\alpha}/\Gamma(1-\alpha)$.
Here, $b>1$ is a dilation (scaling) parameter and $C_{\alpha}(b)$ is a fitting dimensionless constant. 
Physical meaning of $\nu_0$ is a 
high-frequency (short-time) cutoff of the stochastic process $\xi(t)$, $t_l=\nu_0^{-1}$. 
The low-frequency (long-time) cutoff corresponds to $t_h=b^{N-1}t_l$.  
Similar scaling and approximation to a power law are well known in the theory of anomalous relaxation \cite{Hughes,Palmer}.
By adjusting $b$ and $N$ one can approximate power law over about $r=N\log_{10}b-2$ time decades. 
Weak dependence of $r$ on $b$ and $N$ ensures
a very powerful and accurate numerical approach to FLE dynamics. Just a handful of 
auxiliary particles can suffice for all
practical purposes.
Of course, such a Markovian embedding of FLE is not unique \cite{Kupferman}. However, 
it appeals by a clear physical interpretation. 
Recently, the set (\ref{embedding1}) was formally generalized to numerically integrate also
superdiffusive FLE \cite{SiegleEPL}.

\section{Flashing ratchet}

To study flashing ratchet transport we take $f(x,t)=f_0(x)r(t)$. Here, $f_0(x)=-U'(x)$ is a deterministic force
acting on the particle (Brownian motor) when a potential $U(x)$ is switched on.  Furthermore,  $r(t)$ switches on/off 
taking  just two values, zero and one, $r(t)=\{1,0\}$. For $U(x)$ we take a typical ratchet potential 
\begin{eqnarray}
\label{U(x)}
U(x)=-U_0\left[\sin\left(\frac{2\pi x}{L}\right)+\frac{1}{4}\sin\left(\frac{4\pi x}{L}\right)\right],
\end{eqnarray}
with amplitude $U_0$ and period $L$. For $r(t)$ two models are considered: 
periodic \textit{vs.} random switching.
In the periodic case, $r(t)=1/2\left[1+{\rm sign}\left(\sin(2\pi\nu t)\right)\right]$, 
where ${\rm sign}(\cdot)$ is standard signum function and $\nu$ is linear frequency of oscillations. The first switching occurs 
at $t_{\rm I}=1/2\nu$ and the total number of switching is $N_{sw}\equiv T_{tot}/t_{\rm I}=2\nu T_{tot}$, where $T_{tot}$ is a total 
computation time which equals $N_{s}dt$; $N_{s}$ is a total number of integration steps of Eq.(\ref{GLE}), 
$dt$ is a time-step. Finally, one can write $N_{sw}=2\nu N_{s}dt$. For random switching $r(t)$, we consider the
model of dichotomous Markovian process (DMP) with the probability $p=2\nu dt$ to switch within $dt$. Such DMP is symmetric
with equal residence time distributions in two states, $\psi(\tau)=2\nu\exp(-2\nu \tau)$. Here,  $2\nu$ is 
transition rate from one state to another one (and also inverse of the mean 
residence time in a state). The mean flashing frequency $\langle \nu\rangle $ equals $\nu$.

In all simulations we scale coordinate in units of $L$ and 
time in units of $\tau_v=(m/\eta_{\alpha})^{1/(2-\alpha)}$. 
It is assumed to be temperature-independent in accordance with the underlying Hamiltonian model \cite{GLE2,WeissBook}. 
This is a standard assumption done also in other ratchet models \cite{ReimannRev,MarchesoniRev}.
Furthermore, energy is scaled in $E=L^2\eta_\alpha^{2/(2-\alpha)}/m^{\alpha/(2-\alpha)}$ and temperature $T= E/k_B$. 
In this work we choose $\alpha=0.5$ and $b=10$, $C_{1/2}(10)=1.3$,  and use
$\nu_0=100$, $N=12$. Such a set of parameters 
gives an excellent approximation to the FLE dynamics over at least ten time decades, until $t_{\rm max}=10^8$.
Similar to \cite{G09,G10} this was checked by comparison with the exact analytical solution 
for the position variance obtained within GLE and FLE \cite{GLE,FLE} in the force-free case. The numerical errors
due to the memory kernel approximation are negligible as compare with the overall statistical error in our 
stochastic simulations (several percents), see for details in \cite{G09}. Like in \cite{arxiv}, numerical solutions
of stochastic differential equations were performed on the graphical processor units (GPUs) with double precision using
stochastic Heun method. The use of GPU computing allowed to parallelize and accelerate simulations by a factor of about 100, 
as compare with conventional CPU computing. 

Schematically, viscoelastic dynamics in the flashing ratchet potential is shown in 
figure \ref{Fig1} (see also video abstract).
The auxiliary particles  are placed on the level $U=0$ and they are not influenced by potential.  
Initially, when the potential is switched on, the Brownian particle moves in the potential well (see snapshots a and b). 
When the potential is off (snapshots c and d) the Brownian particle  ``freely'' subdiffuses, interacting only with
auxiliary particles. After the potential is switched on again, Brownian particle moves into the next potential well 
(see snapshot e) with the net motion occurring in the negative direction. During such motion not all auxiliary particles 
follow immediately to the Brownian particle. 
Some of them are at large distances from the Brownian particle 
and are essentially less mobile than the Brownian particle and its nearest environment 
because they have essentially larger
friction coefficient than other auxiliary particles. The weaker is the coupling constant $k_i$ of 
an auxiliary
particle to the Brownian one the larger is the corresponding frictional coefficient $\eta_i$.  
These very sluggish particles  create a slowly fluctuating quasi-elastic force (a temporal biasing force 
\cite{G09,G12}) acting on the central particle.

\section{Results and discussion}

In all simulations we  monitored two main statistical quantities, the mean position 
$\langle x(t)\rangle$ and the position variance
$\langle \delta x^2(t)\rangle=\langle x^2(t)\rangle-\langle x(t)\rangle^2$. 
For the ensemble average, $N_{\rm ens}=10^4$ trajectories were used.
The focus is on such physical 
quantities as subdiffusive current (subvelocity) $v_\alpha$, subdiffusion coefficient $D_\alpha$ and  
generalized Peclet number\cite{G10} ${\rm Pe}_{\alpha}:=v_{\alpha}/D_{\alpha}$. The latter one surves 
as a measure for the coherence and 
quality of anomalous stochastic transport, by analogy with normal one \cite{Lindner}. 
The subvelocity and the subdiffusion coefficient are defined as usually \cite{Metzler,G10,GH11,G12}
\begin{eqnarray}
 v_\alpha=\Gamma(1+\alpha)\lim\limits_{t\to\infty}\frac{\langle x(t)\rangle}{t^\alpha},\\
 D_\alpha=\frac{1}{2}\Gamma(1+\alpha)\lim\limits_{t\to\infty}\frac{\langle\delta x^2(t)\rangle}{t^\alpha}\;.
\end{eqnarray}
Here, the limit is understood in the following sense: $t$ is large but still much smaller than the memory cutoff, which 
can always be made unreachable in numerical simulations and thus is irrelevant for the results presented. 
Practically, the
values of $v_\alpha$ and $D_\alpha$ were calculated by fitting the numerical 
dependencies $\langle x(t)\rangle$ and 
$\langle\delta x^2(t)\rangle$ with a power-law function
 $a_{\{v,D\}}t^\alpha$, extracting the corresponding $a_v$ and $a_D$ 
within the last time window of simulations. 
Total time of simulations was varied in the interval $[2\times10^5, 10^6]$ depending on system parameters 
to guarantee a good fit with $\alpha=0.5$ (convergency
is slow); 
the time step was fixed at $dt=2\times10^{-3}$.

\begin{figure}
 \centering
 a)\includegraphics[height=56mm]{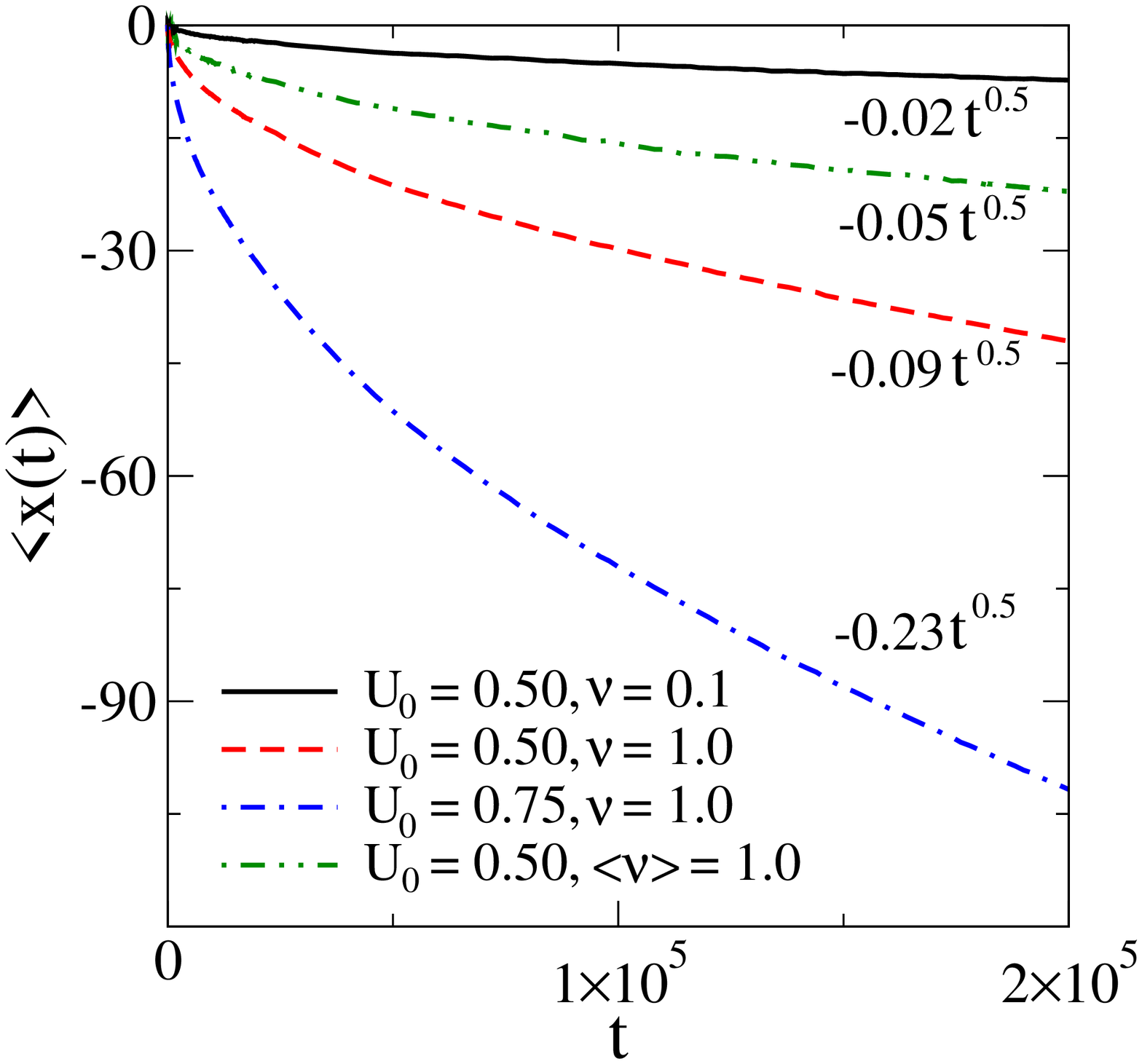}\hspace{5mm}
 b)\includegraphics[height=56mm]{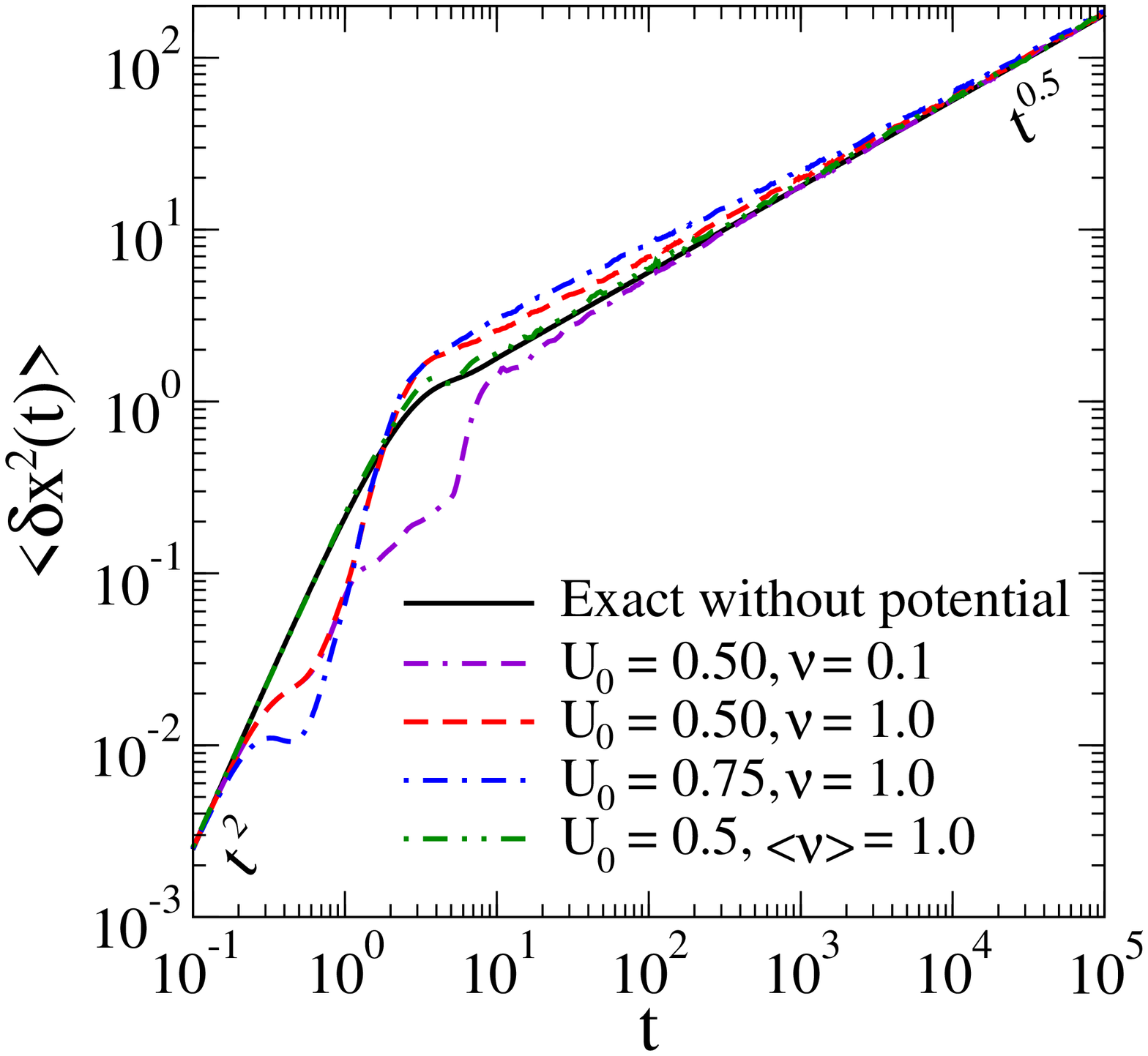}
 \caption{The mean particle position (a) and mean variance (b) at different values of the potential 
height $U_0$ and flashing frequency $\nu$ for periodic and random (green dash-dot-dot line) driving 
at $T=0.25$. The black line in (b) depicts exact analytical FLE result for the potential-free case. }
 \label{Fig2} 
\end{figure}

Typical dependencies for the mean particle position and variance on time for some
different values of the potential amplitude  $U_0$ and flashing frequency  $\nu$ are shown 
in figure \ref{Fig2}(a) and (b), respectively. 
Here black solid, red dash and blue dash-dot lines correspond to the periodic flashing, 
whereas green dash-dot-dot line relates to the random one. The direction 
of transport in figure \ref{Fig2}(a) is negative and the transport has clearly subdiffusive character. 
We indicate in figure \ref{Fig2}(a) also the corresponding power-law asymptotics. 
One can notice that stochastic flashing tends to delay the corresponding transport process in  
comparison with the periodic one of the same frequency (cf. red dash \textit{vs.} green dash-dot-dot lines),
and the transport becomes faster with the increased flashing frequency (this is but not a universal feature, see
below). The diffusion is 
initially always ballistic (universal regime), 
$\langle\delta x^2(t)\rangle=v_T^2t^{2}$, where $v_T=\sqrt{k_BT/m}$ is thermal velocity. This is because
the Brownian particle's velocity is initially thermally distributed and the action of the medium requires 
some time 
to settle in. After some transient, subdiffusion follows asymptotically to one and the 
same universal dependence,
$\langle\delta x^2(t)\rangle = 2D_{\alpha} t^{\alpha}/\Gamma(1+\alpha)$ with $D_{\alpha}=k_B T/\eta_{\alpha}$,
as in the absence of potential, independently of the potential height and the presence of driving, cf. in 
figure \ref{Fig2}(b). 
This universality (or a weak sensitivity in the case of a strong fast
driving \cite{G10}) is a benchmark of viscoelastic subdiffusion \cite{G09,G10,GH11,G12}. 
We  elaborate on this fact in more detail  below.

\subsection{Ergodicity}

\begin{figure}
 \centering
 a)\includegraphics[height=56mm]{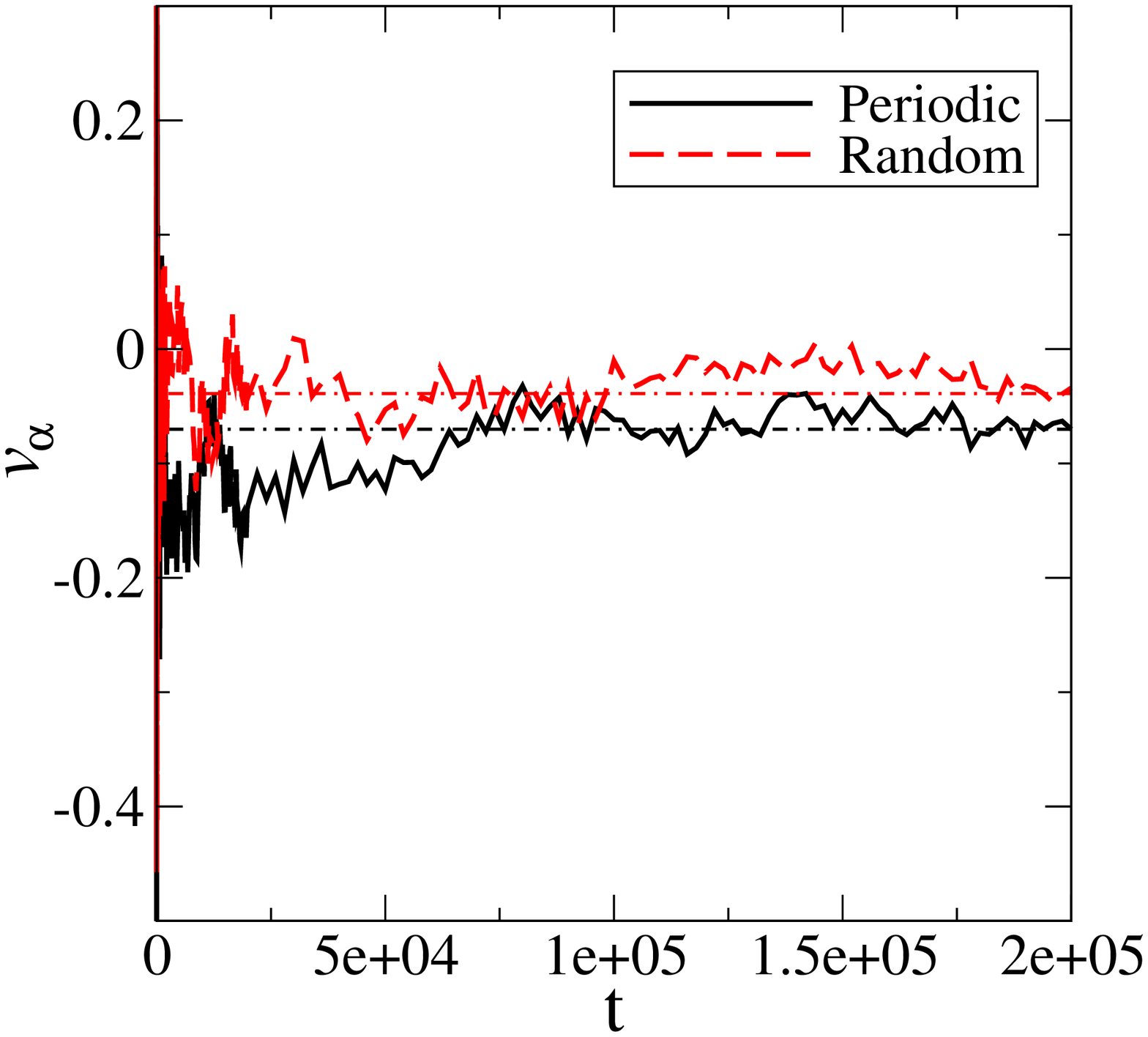}\hspace{5mm}
 b)\includegraphics[height=56mm]{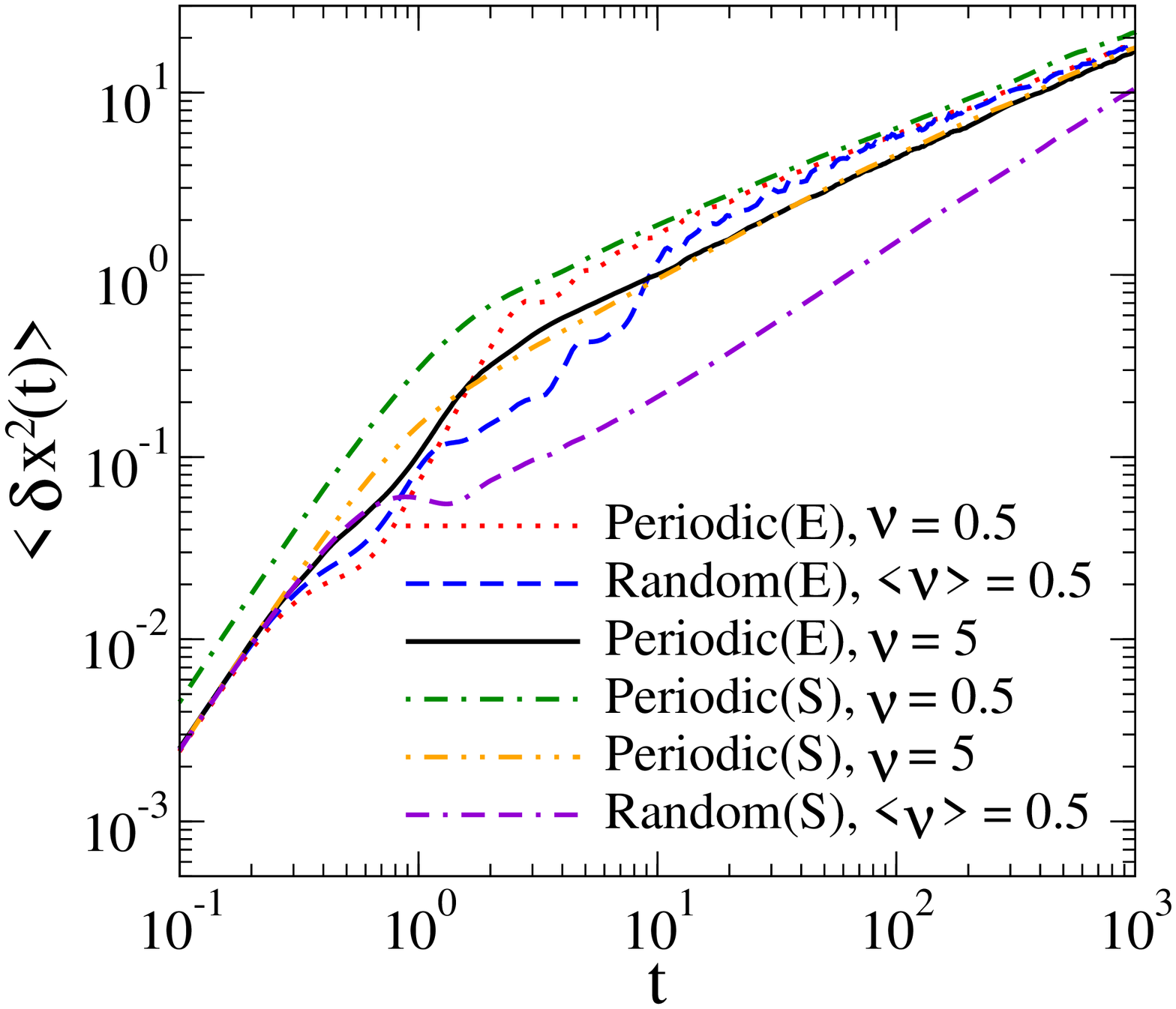}
 \caption{Subvelocity $v_\alpha(t)$, obtained for the single trajectory a) for periodic and random flashing ratchets 
 at $U_0=0.5$, $T=0.25$ 
and linear frequency of flashing (its mean value) $\nu=0.5$. Mean deviation $\langle\delta x^2(t)\rangle$ b) 
for ensemble (E) and single trajectory (S)
averaging at $U_0=0.5$ and $T=0.25$ for periodic and random flashing ratchets. 
The time averaging for a single trajectory is done using 16-exponential approximation to memory kernel and 
${\cal T}=2\times 10^7$. }
 \label{Fig3} 
\end{figure}

An important issue in anomalous transport is ergodicity \cite{Yaglom,Papoulis}, i.e. whether time-average over a single particle
trajectory delivers non-random result, same for all identical particles, or a principal randomness and scatter in the single-particle
averages emerge. If this is the case, then
the behavior
and fate of each individual though identical particle is different, even in the limit of infinitely long trajectories, 
and only ensemble-averaging
smears out this principal randomness and delivers non-random result. In the case of ergodic transport, 
a single-trajectory average should coincide with the
result of the ensemble averaging. For example, CTRW subdiffusion with divergent
mean residence times in traps is patently nonergodic \cite{Bel}
and single-trajectory averages obey some universal fluctuation laws \cite{Bel,He,Lubelski,Sokolov} leading to 
a universal scaling relating such transport and diffusion in periodic
potentials \cite{GH06}. On the contrary, viscoelastic subdiffusion, both free 
\cite{Deng}, and in time-independent potentials \cite{G09} was shown to be mostly ergodic, though some transient
nonergodic features can be present. Whether it remains ergodic also in time-fluctuating fields presents a nontrivial
problem. First, we checked the asymptotic ergodicity of ratchet viscoelastic transport
which is expected.  To show that this indeed is the case we plotted the 
fluctuating subvelocity $v_\alpha(t)$, obtained for a single trajectory 
as $v_\alpha(t)=\Gamma(1+\alpha)x(t)/t^\alpha$ 
in figure \ref{Fig3}(a)  \textit{vs.} the ensemble-average obtained with $10^4$ particles. One can see that $v_\alpha(t)$
fluctuates around the ensemble average with the amplitude of fluctuations diminishing with $t$. In the limit
$t\to\infty$ the both averages coincide and the transport is ergodic. 
For any finite $t$, statistical fluctuations are present, of course, also in
ergodic case. Next, a single trajectory average of the  squared displacement can be obtained as \cite{He,Lubelski,Deng,G09},
\begin{eqnarray}\label{time-av}
\langle \delta x^2(t)\rangle_{\cal T}=\frac{1}{{\cal T}-t}\int_{0}^{{\cal T}-t}[x(t+t')-x(t')]^2dt'.
\end{eqnarray}
 In the ergodic case,
this average should coincide in the limit
${\cal T}\to \infty$ but for any $t$ with the result of the ensemble averaging.
If the agreement holds only for large $t$, then diffusion is asymptotically ergodic. For viscoelastic subdiffusion
in fixed periodic potentials the agreement holds both for small $t$ (ballistic regime) and for large $t$ \cite{G09}. However, some
deviations from ergodicity occur for intermediate $t$, on the time scale of diffusion over several potential periods.
This is because even if the mean escape time exists the escape kinetics remains anomalous  for activation  
barriers of an intermediate height \cite{G09}.

Diffusional  spreading derived from single trajectories is compared with the corresponding ensemble averages
in figure \ref{Fig3}(b). 
In the ballistic regime, using $x(t+t')-x(t')\approx v(t')t$ for small $t$
in Eq. (\ref{time-av}), one obtains 
\begin{eqnarray}\label{time-av1}
\langle \delta x^2(t)\rangle_{\cal T}\approx t^2\frac{1}{{\cal T}-t}\int_{0}^{{\cal T}-t}v^2(t')dt'=
\langle \delta v^2(t)\rangle_{\cal T} t^2.
\end{eqnarray}
Clearly, if  in the limit ${\cal T}\to\infty$ the time average of $v^2(t)$ coincides  with the 
ensemble average yielding $v_T^2$,  then ergodicity  holds also in the ballistic regime. In the absence of
driving this is indeed the case \cite{G09}. However, a periodic driving exciting coherent oscillations emerging
due to a combined action of the external trapping potential and the viscoelastic cage effect 
can cause large additional temporal fluctuations in $v(t)$ which are not self-averaged in $\langle v^2(t)\rangle_{\cal T}$.
Accordingly, such a periodically driven subdiffusion ceases to be ergodic in the ballistic regime. 
Temporally but on a large intermediate time scale ergodicity is generally broken.
Nevertheless, asymptotically
it remains ergodic as figure \ref{Fig3}(b) implies. Convergency to such asymptotically ergodic regime is very slow.
However, it improves with the increased frequency of 
flashing, see the case $\nu=5$ in figure \ref{Fig3}(b). For random driving, ergodicity in the discussed sense holds also
in ballistic regime.

\subsection{Frequency dependence}

\begin{figure}
 \centering
 \includegraphics[height=56mm]{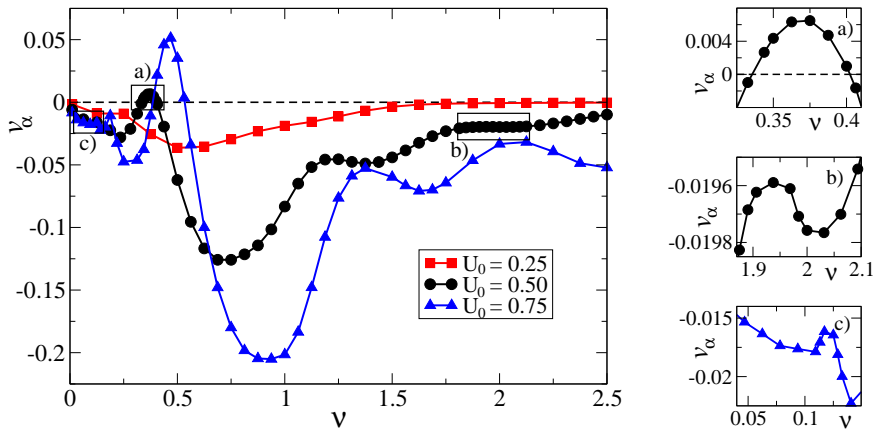}
 \caption{Anomalous current (subvelocity $v_\alpha$) as a function of linear frequency of periodic flashing at 
 fixed temperature $T=0.25$ for different values $U_0$. }
 \label{Fig4} 
\end{figure}

\begin{figure}
 \centering
 \includegraphics[width=80mm]{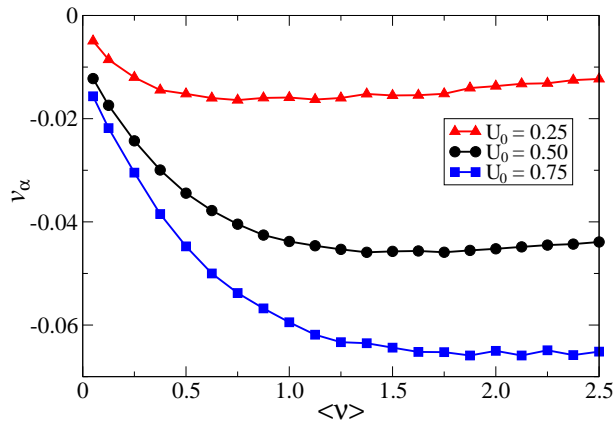}
 \caption{Anomalous current (subvelocity $v_\alpha$) as a function of mean linear frequency of random flashing 
 at fixed temperature $T=0.25$ for different values $U_0$. }
 \label{Fig5} 
\end{figure}

We proceed further with the dependence of anomalous current $v_\alpha$ on frequency $\nu$ at a fixed value 
of temperature for different potential amplitudes $U_0$. In the case of fluctuating tilt subdiffusive ratchets \cite{G10}
this dependence was especially intriguing. That anomalous transport vanished in the limit of vanishing modulation frequency (adiabatic
driving limit) and exhibited a maximum at intermediate frequencies.  The origin of this maximum was related to 
stochastic resonance effect in Ref. \cite{arxiv}.
This is in a sharp contrast with normal diffusion fluctuating tilt ratchets where transport optimizes namely in the adiabatic limit.    
The results for periodic flashing are shown in figure \ref{Fig4} and for stochastic driving in figure \ref{Fig5}. For small driving frequencies
subdiffusive transport occurs always in the natural negative direction, which is opposite to the natural positive direction of
fluctuating tilt ratchet.  Our anomalous Brownian motors share these features with their normal diffusion 
counterpart \cite{ReimannRev,MarchesoniRev}. 
There are but novel features which are due to a profound role of the inertial effects in our model. We recall that in the used scaling the
unit of time corresponds to a scaling time constant $\tau_v$ of the decay of velocity autocorrelation function. Correspondingly,
the ballistic regime occurs until $t\sim 1$, when the potential is off.  
For periodic driving and a small potential height $U_0=0.25=T$ (see red line with squares in figure \ref{Fig4}) subvelocity 
takes negative values for all  flashing frequencies considered. The dependence of $|v_\alpha|$ on $\nu$ has a unique 
maximum at $\nu_{c0}\simeq 0.5$, where subdiffusive transport is optimized, with negative $v_{\alpha}$. 
For the fast-flashing regime, when $\nu>1$ 
subvelocity asymptotically decays to zero with $\nu$. Increase in the ratchet potential height $2U_0$ accompanied by an increased role
of the inertial effects leads to a profound 
change in the character of the frequency-dependence $v_\alpha(\nu)$ (see black curve with circles for $U_0=0.5$ in figure 
\ref{Fig4}). Let us study 
this dependence in more detail. The first difference from the former case is a multiple (double) current inversion, 
realized around  $\nu_{c1}\simeq 0.335$ and $\nu_{c2}\simeq 0.402$. 
Enlarged picture of this inversion effect is shown 
in the insert $a)$ placed on the right panel in figure \ref{Fig4}. Here, a well defined peak occurs at  $\nu_{\max 1}=0.375$
with positive values 
for $\nu_{c1}<\nu<\nu_{c2}$. Furthermore, on the left side of  $\nu<\nu_{c1}$ there is a maximum of $|v_{\alpha}|$
for negative $v_{\alpha}$. Increase of $\nu$ beyond $\nu_{c2}$ leads to a global maximum of 
$|v_\alpha|$  realized at $\nu_{\max 2}\simeq 0.75$. With a further increase in flashing 
frequency a complicated oscillatory pattern with several more relative minima and maxima in $|v_{\alpha}|$
emerges. With a further increase of the potential amplitude these oscillatory resonance-like features become more pronounced --
compare two cases with $U_0=0.50$ and $U_0=0.75$ in figure \ref{Fig4}. This is clearly related to an increased role of the inertial effects.
First, the amplitude of oscillations in $v_{\alpha}(\nu)$ increases. Second, ever more extrema appear, see insets b) and c)
in figure \ref{Fig4}.

The situation is quite different in the case of random flashing. Dependencies of anomalous current 
versus mean linear frequency $\langle\nu\rangle$ at fixed temperature $T=0.25$ for different values of $U_0$ 
are shown in figure \ref{Fig5}.
Here, contrary to the case of periodic flashing one has more simple dependencies $v_\alpha(\langle\nu\rangle)$ with 
one broad minimum for each considered potential amplitude $U_0$. Increase in $U_0$ leads to increase in absolute 
values for subvelocity, broadening the minimum and shifting minimum position towards larger values of mean 
linear frequency. Thus, 
in the random flashing case the system is not manifestly sensitive to inertial effects due to stochastic 
nature of flashing. The maximal value of subvelocity becomes also much smaller, 
compare figure \ref{Fig4} and figure \ref{Fig5}, due to the absence of resonances.

\begin{figure}
 \centering
 \includegraphics[width=80mm]{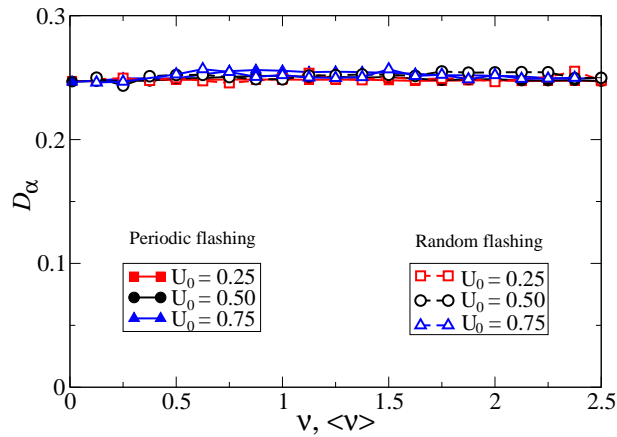}
 \caption{Sub-diffusion coefficient $D_\alpha$ as a function of linear frequency $\nu$ 
 (or mean linear frequency $\langle\nu\rangle$) for periodic and random flashing, respectively, at 
 fixed temperature $T=0.25$ and different values $U_0$. }
 \label{Fig6} 
\end{figure}

A benchmark feature of viscoelastic subdiffusion is that it asymptotically does  not depend on the presence
of static periodic potential \cite{G09,GH11,G12}. It was also only weakly dependent on driving in the case
of rocking subdiffusive ratchets \cite{G10}. The diffusional spread in flashing ratchets is also practically
not dependent on the potential amplitude $U_0$ and flashing frequency as figure \ref{Fig6} illustrates.  
One can see, that for both periodic and random flashing the anomalous diffusion coefficient does 
not display any profound dependence on the flashing frequency and the potential amplitude. It 
takes values around $0.25$, which is $D_{\alpha}=T$ in the scaled units,  with a deviation which is
less than 3\% and lies within the statistical error margin of our simulations. 
Therefore, the corresponding generalized Peclet number ${\rm Pe}_{\alpha}:=v_{\alpha}/D_{\alpha}$, which measures 
the coherence and quality of transport, resembles the behavior of the absolute value of subcurrent in 
figures \ref{Fig4} and \ref{Fig5}, which should be multiplied by factor 4 in order to obtain  
${\rm Pe}_{\alpha}$. 

\subsection{Temperature dependence}

\begin{figure}
 \centering
 a)\includegraphics[height=56mm]{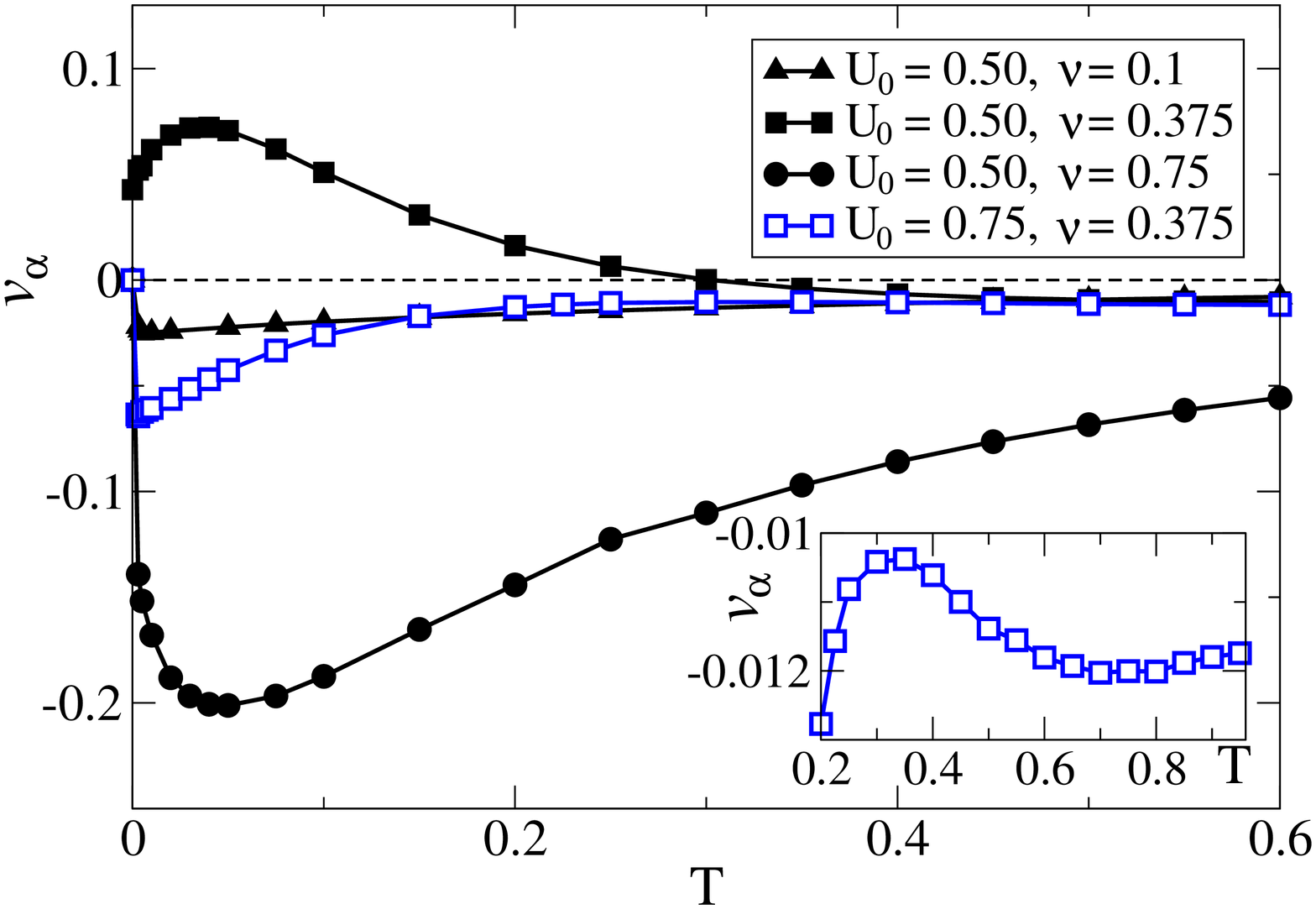}\hspace{5mm}
 b)\includegraphics[height=56mm]{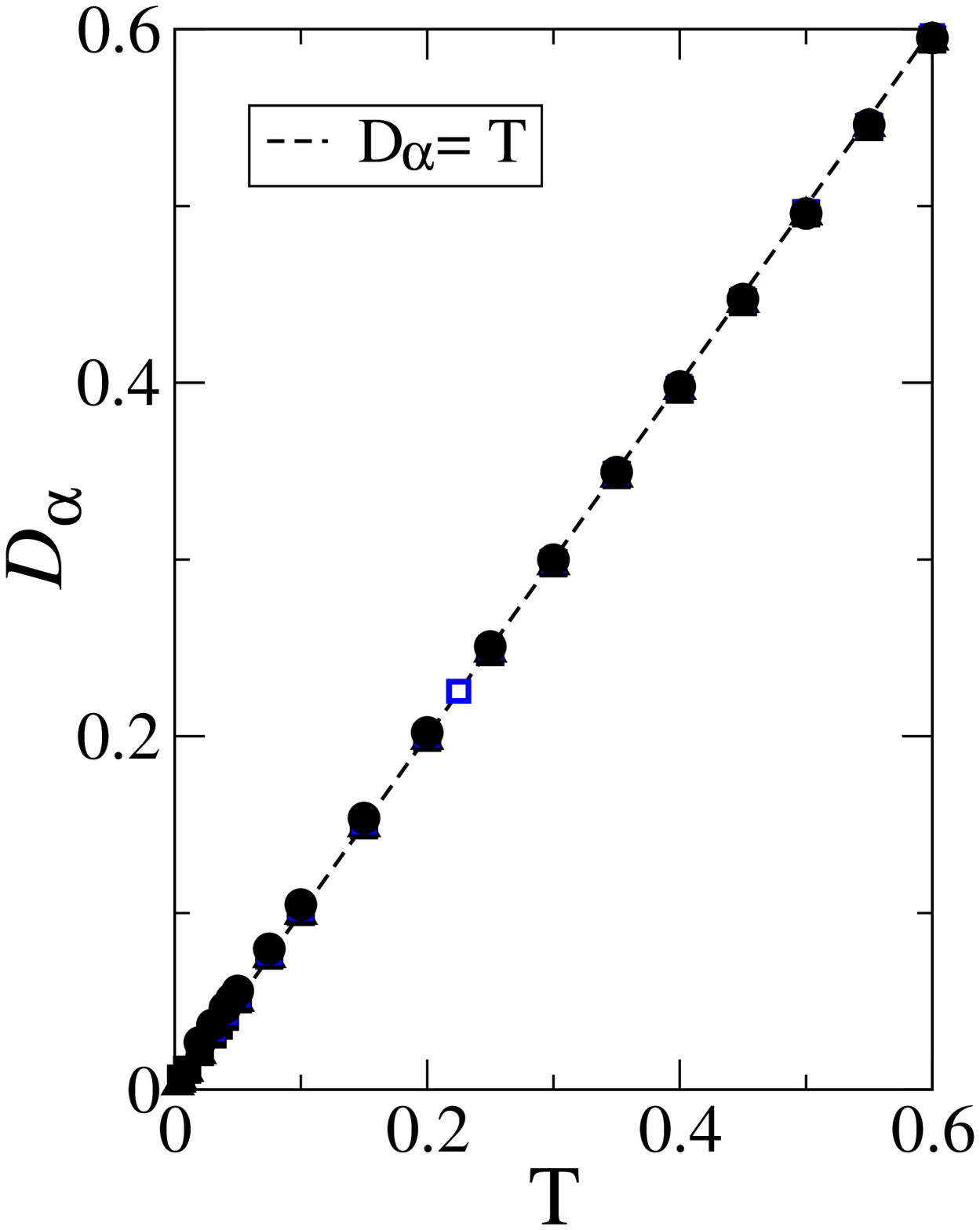}\\ \vspace{5mm}
 c)\includegraphics[height=56mm]{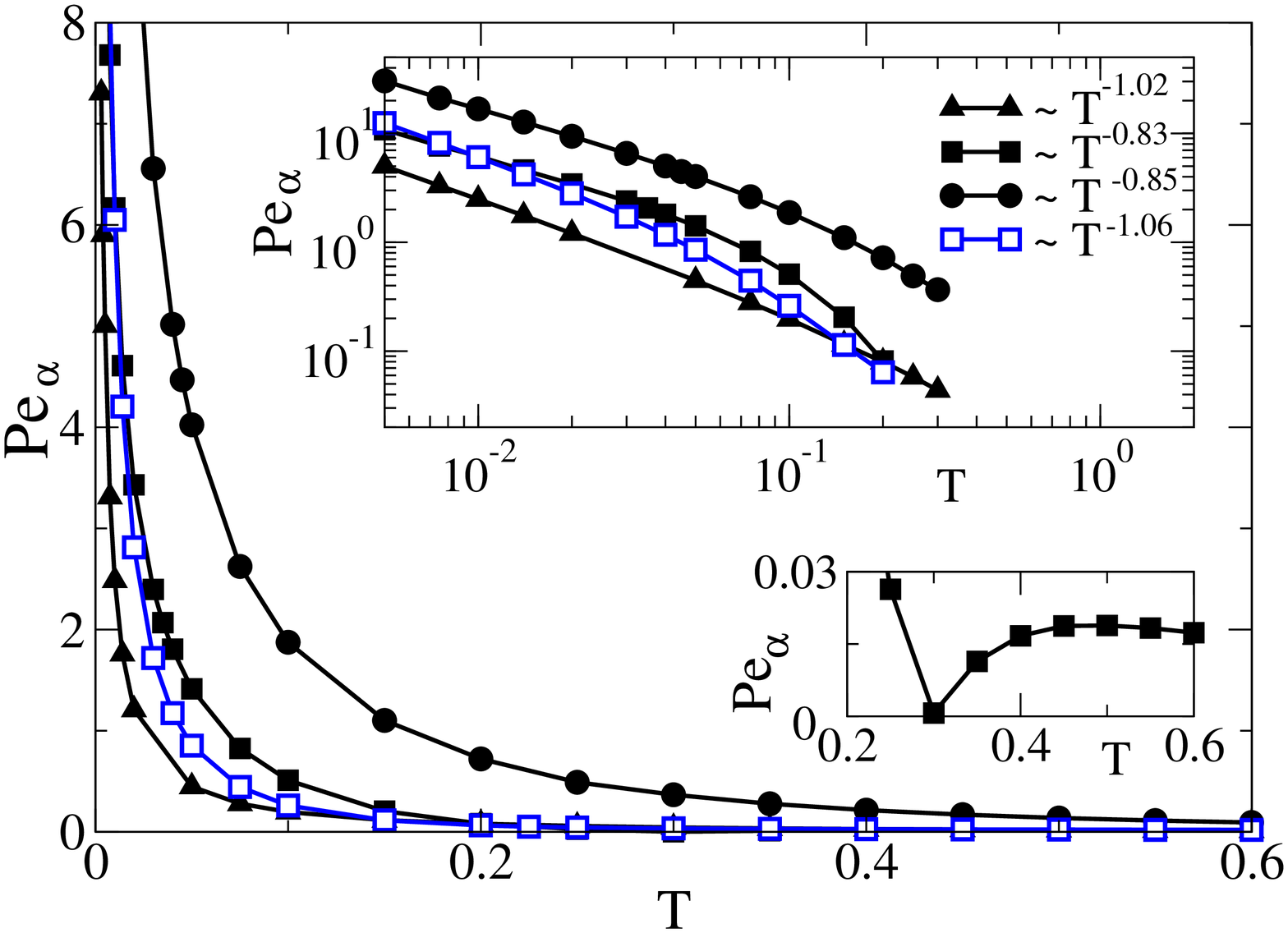}
 \caption{Anomalous current (subvelocity) $v_\alpha$ (a), subdiffusion coefficient $D_\alpha$ (b) and generalized 
 Peclet number ${\rm Pe}_{\alpha}$ (c) versus temperature for the periodic flashing ratchets at different values 
 for linear frequency of flashing $\nu$ and potential height $U_0$. The legend for (a), (b) and (c) figures is the same.}
 \label{Fig7} 
\end{figure}

\begin{figure}
 \centering
 a)\includegraphics[height=56mm]{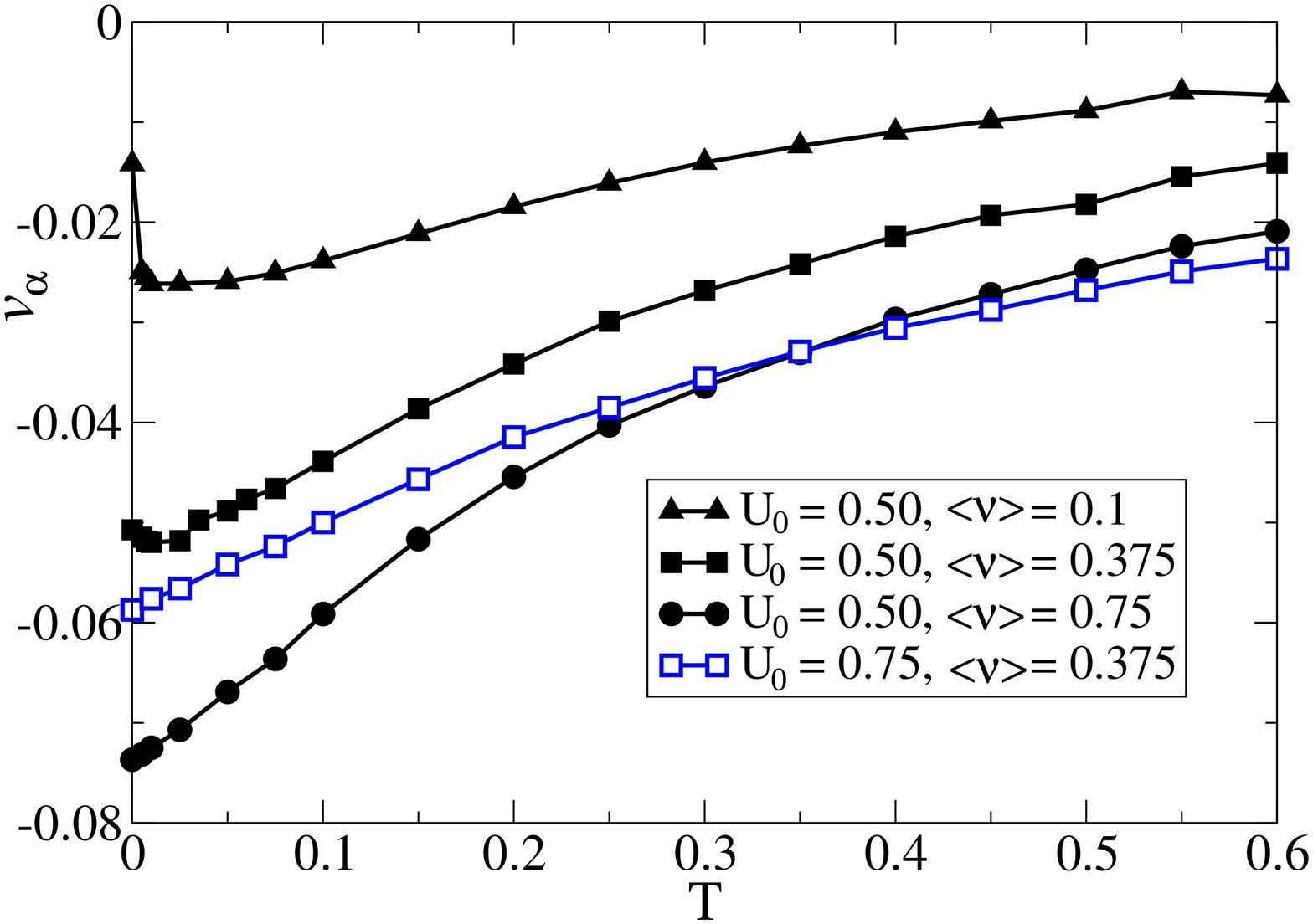}\hspace{5mm}
 b)\includegraphics[height=56mm]{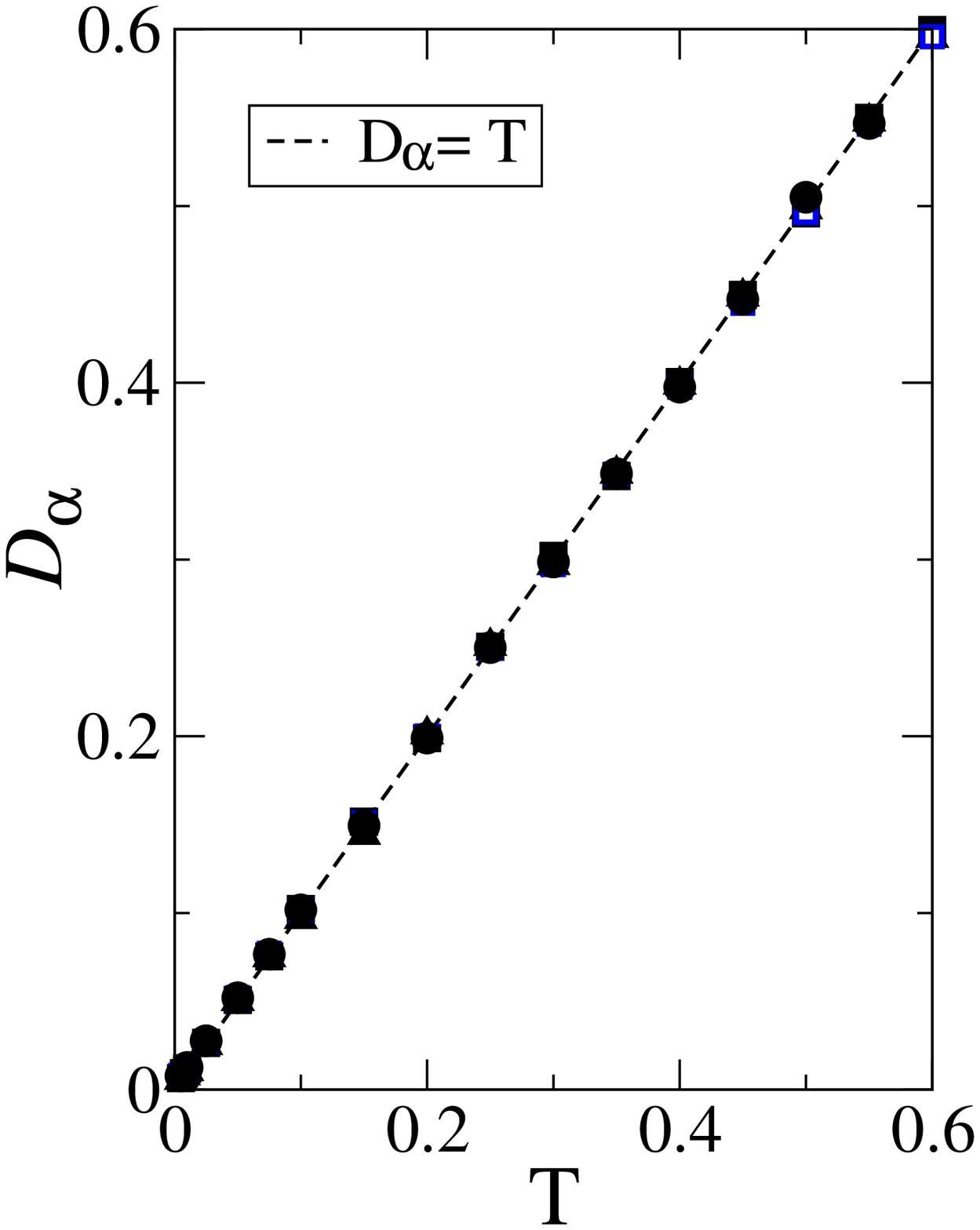}\\ \vspace{5mm}
 c)\includegraphics[height=56mm]{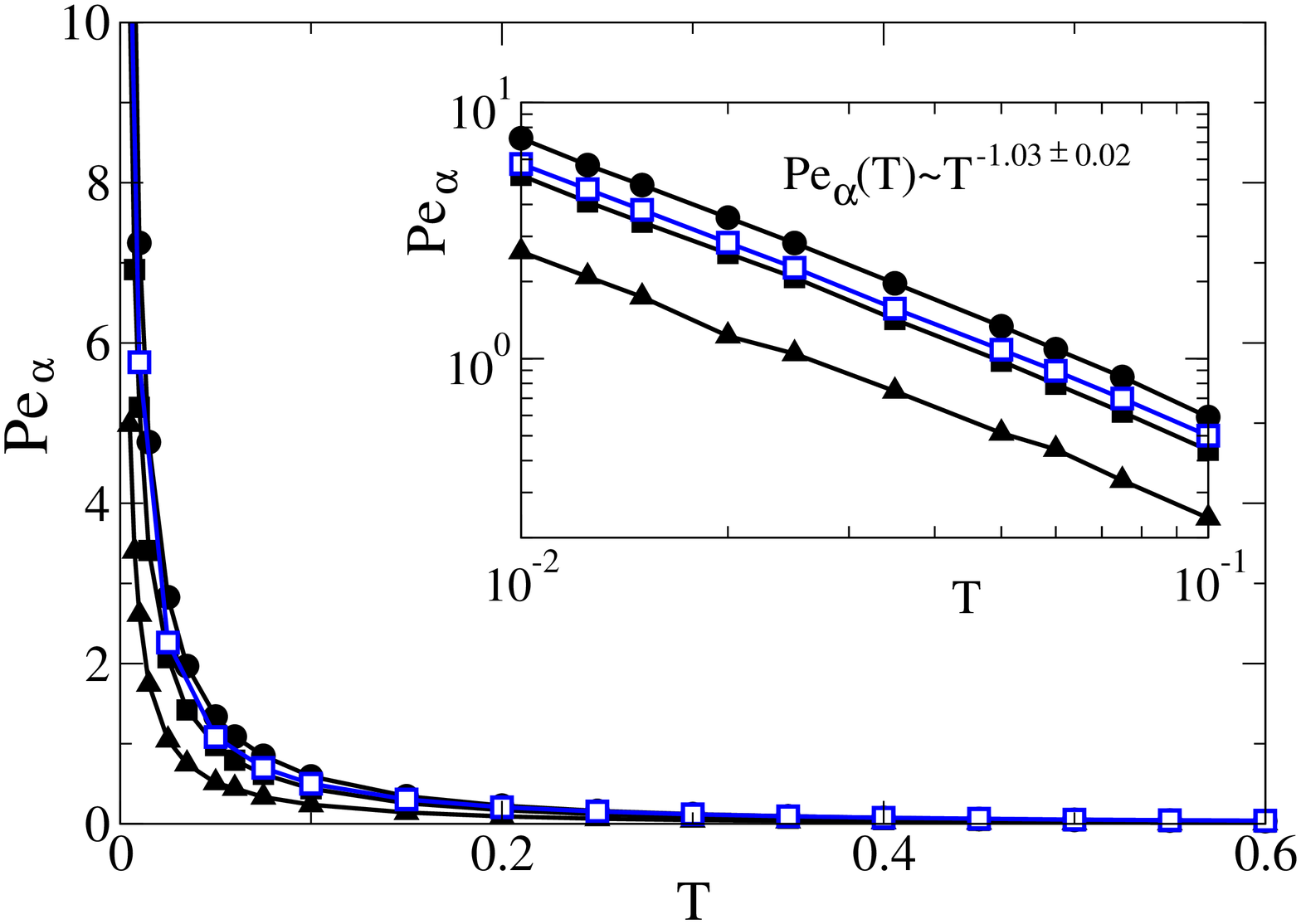}
 \caption{Anomalous current (subvelocity) $v_\alpha$ (a), subdiffusion coefficient $D_\alpha$ (b) and generalized 
 Peclet number ${\rm Pe}_{\alpha}$ (c) versus temperature for the random flashing ratchets at different values 
 for mean linear frequency of flashing $\langle\nu\rangle$ and potential height $U_0$. 
 The legend for (a), (b) and (c) figures is the same.}
 \label{Fig8} 
\end{figure}

The temperature dependence of subvelocity $v_\alpha$, and generalized Peclet number ${\rm Pe}_\alpha$
is mostly intriguing, whereas the dependence of  subdiffusion coefficient $D_\alpha$ on temperature is expected to
be trivial, $D_\alpha(T)=T$. We plotted numerical results for all these three quantities in figure \ref{Fig7} (periodic
flashing) and  figure \ref{Fig8} (stochastic flashing) for three different test values of the flashing frequency,
$\nu=0.1$ (low-frequency),  $\nu=0.375$ (intermediate frequency), and $\nu=0.75$ (high-frequency).
The temperature dependencies of subvelocity share two common features, see figures \ref{Fig7}(a)
and \ref{Fig8}(a). First, the subvelocity
is finite in the limit $T\to 0$ (this is always so for random flashing, but only for certain frequency
windows in the case of periodic flashing). 
Second, it diminishes to zero in the limit of high temperatures. The latter  is expected,
but the former is not, being highly surprising. Indeed, in a combination with the subdiffusion coefficient following to
the expected linear dependence and vanishing in the limit $T\to 0$ [see figures \ref{Fig7}(b)
and \ref{Fig8}(b)], the non-zero value of $v_{\alpha}$ in this limit means that the basic mechanism of 
operating our subdiffusive viscoelastic Brownian motors is very different from one in the case of normal Brownian
motion. This is at odds with our expectations (see Introduction). 
Indeed, in the absence of viscoelastic effects 
the flashing Brownian motors of the studied kind are known to be essentially 
based on thermal diffusion, i.e.
such transport is assisted by thermal fluctuations \cite{NelsonBook}. It is therefore 
expected to vanish in the limit $T\to 0$, where the thermal fluctuations vanish (in neglecting
quantum effects). 
Our results say, however, that
viscoelastic flashing ratchets are not based primarily on the thermal diffusion and  thermal fluctuations.  
These are long-range elastic correlations
which are at work here in combination with the potential fluctuations.  
Namely, the external field acting on the Brownian particle changes abruptly (on-off process)
the energy stored in the elastic springs. Not only the Brownian particle moves under the influence of $f_0(x)$, but also 
the auxiliary particles under the influence of Brownian particle. Some of them rapidly adjust their positions,
but there are also slow particles which cannot follow immediately. When the potential is off these extra sluggish
particles pull the Brownian 
particle. This is why its motion is generally not frozen even at $T=0$ within our purely classical setup. 
Because subdiffusion
diminishes with lowering temperature, the generalized Peclet number grows accordingly, ${\rm Pe}_{\alpha}(T)\sim 1/T^\delta$
with $\delta$ in the range $[0.83,1.05]$, see figure \ref{Fig7}(c), and figure \ref{Fig8}(c). Clearly, ${\rm Pe}_{\alpha}\to\infty$
in the limit $T\to 0$, where the transport becomes perfect.

There are also profound differences in the temperature-dependence of anomalous transport in the cases of periodic and random
flashing. So, in the case of periodic flashing the transport can occur in the counterintuitive positive direction, see
the curve for $U_0=0.5$, $\nu=0.375$ in figure \ref{Fig7}(a). In this case, $v_{\alpha}$ remains finite also at $T=0$, and an
inversion of the current direction with increasing temperature is possible.
This is not so for other three curves in figure \ref{Fig7}(a), where transport vanishes jump-like at $T=0$. 
Therefore, it seems obvious that for a periodic
flashing there are frequency windows for permitted transport at $T=0$ (a detailed investigation
of this issue is left for a separate study). On the contrary, for  random driving in  
figure \ref{Fig8}, a, the transport remained always finite at $T=0$. Moreover, here transport
always occurs in the intuitive negative direction.  Both in the random case and in the periodic case 
the transport can be optimized
at some temperature $T_{\rm max}\neq 0$. It can be but also that $T_{\rm max}=0$.

\section{Conclusions}

With this work we put forward anomalously slow Brownian motors based on flushing subdiffusion in viscoelastic media. 
Both anomalous transport and subdiffusion  were shown to be asymptotically ergodic.  However, such driven 
subdiffusion can
be transiently nonergodic on appreciably long time scales. 
The transport exhibits a remarkably good quality
at low temperatures. It remains finite even at zero temperature in many cases (but not always in the case
of periodic flashing) with zero dispersion. In this limit, the transport is clearly induced by long-range
elastic correlations in a combination with potential flashing and not by the thermal noise of  environment.
This is surprising and at odds with a popular explanation of the origin of noise-induced 
flashing transport in the case memory-free stochastic dynamics \cite{ReimannRev,MarchesoniRev}. 
Moreover, in the case of periodic driving
anomalous current exhibits multiple resonance-like features, can flow in the counter-intuitive direction, and invert its direction
both with the change of flashing frequency and with change of temperature. 
The inertial cage effects are essential for many observed features.
In particular, they are crucial for the resonance-like features manifested for high potential barriers
in the case of periodic flashing.
We hope that our research will stimulate further cross-fertilization of the field of fluctuation-induced transport 
and the field of anomalously slow transport and subdiffusion which flourish at present with little interaction. 
There emerges an increasing experimental
support for the occurrence of subdiffusion in cytoplasm of biological cells and we believe that there might be also place 
for operating sluggish molecular motors of the kind described.

\section{Acknowledgments}

Support of this research by the Deutsche Forschungsgemeinschaft, Grant
GO 2052/1-1 is gratefully acknowledged.

\end{document}